\documentclass[amsmath,amssymb,lengthcheck,aps,prl] {revtex4}
\usepackage[T1]{fontenc}
\usepackage[utf8]{inputenc}
\usepackage{graphics}
\usepackage{graphicx}
\usepackage{xcolor}

\begin{document}

  \title{Low-dimensional Bose-Bose Mixture in Random Speckle Potential}
\author{Avra Banerjee}
\author{Saswata Sahu}
\author{ Dwipesh Majumder }
\affiliation{Department of Physics, Indian Institute of Engineering Science and Technology, Shibpur, WB, India}

\begin{abstract}
In this work, we have studied the effect of the repulsive speckle potential in a mixture of Bose-Einstein condensates in one dimension   (1D) and two dimension (2D). We simulated linear and circular random speckle potentials in 1D and 2D, respectively. Our calculation shows that the condensate density forms a sharp ring in 2D, and the condensate is divided into two parts in 1D at a high impurity density of speckle potential. We have calculated the energy and chemical potential of the system by solving the Gross-Pitaevskii (GP) equation to see the stability of the condensate. In our study, we have seen that the nature of the impurity response is the same for one-dimensional and two-dimensional quantum droplets. 
\end{abstract}
\maketitle
\section{I. Introduction}

The fundamental structure of Bose-Einstein condensation (BEC) may be represented by the Gross-Pitaevskii (GP) energy function as experimentally low atomic energy and densities are obtained. The s-wave interactions dominate the atom-atom interaction in these situations \cite{bec1,bec2}. One exciting topic in ultracold atomic physics is the creation of quantum droplets. A mixture of two species of BECs is a demanding system for studying quantum droplets \cite{Petrov2015}. For quantum droplet, there is an attractive mean-field interaction between the same species of Bose atoms and a repulsive mean-field interaction between different species of Bose atoms such that the total mean-field interaction is attractive. The quantum droplet forms due to the competition between mean-field attraction and repulsive quantum fluctuation, also known as Lee-Huang-Yang (LHY) correction \cite{LHY1,LHY2,LHY_3}. Experimental study was done in the quantum droplets of the mixture of two species of alkali Bose atoms. Quantum droplet may be a mixture of two species of atoms of an element with different degrees of freedom \cite{Trarruell2018, drop_exp2, Trarruell2018PRL, homo}. Quantum droplets of BEC have been seen in heteronuclear mixtures of atoms where both species are in the lowest hyperfine state  \cite{Rb_Na, Itali2020}. In addition to this system, the quantum droplet has also been observed as  1D  droplet in the asymmetric dipolar interacting systems \cite{dipolar_droplets}. Including three-body interactions, quantum droplets can be formed \cite{3body2}.

Low-dimensional quantum droplets have the potential advantage of a long lifetime due to the reduced Hilbert space and lower probability of three-body collisions. Earlier, it was predicted that condensate could not be formed in one dimension, but later, Bogoliubov's theory predicted that it is possible to create the condensate \cite{bog_1_lhy, bog_2_lhy}. Theoretical study of the liquid form of BEC in the lower dimensions can be found here \cite{2D_liquid,2D_liquid1, 2D_liquid2,2D_liquid4,1D2D2018, 2D_liquid3, Gajda19, low_Gajda2023}. Our motivation is to study the effects of the random speckle potential on a self-bound liquid droplet, and there are many interesting studies with the random disorder. Some of them are listed below. BECs in optical potentials that are effective instruments for quantum simulation of disordered systems  \cite{leonard}. The static and dynamic features of  Bose-Einstein condensates in disorder are investigated using an optical speckle potential \cite{speckle1}. The speckle potential is both controllable and random \cite{speckle1, speckle2, speckle3, speckle4, speckle5}. Many more disordered potentials have been investigated in BEC \cite{Gaussian1, Gaussian_Adhikari, GP_random2, Gaussian2014, Gaussian2018,Lorentzian}. There are multiple domains to random disorder in the Bose system \cite{Anderson2007, Anderson2010, Anderson2008, Anderson2021,SIT3_2004, SIT1_2014, SIT2_2013,BBM-2020,2d_sahu}. 
Impurity potential sometimes enhances confinement, even without external confinement  \cite{boundImp1}.  There are different theoretical methods to study the disordered Bose system, which are finite difference method of solution of the GP equation \cite{GP_random1, GP_random2, GP_random3},  the Monte-Carlo (MC) method \cite{PIM_2010_Pilati,diff_mc,quantum_mc} and the perturbative method \cite{analytical_ran1, MFT1}. 

In this article, we studied quantum droplet's density, energy, and chemical potential variation in the presence of circular (2D case) and linear (1D case) speckle potential. The research relies on extended GP equations. In both cases, we have used reduced single GP equation \cite{2D_liquid, 2D_liquid3, 2D_liquid4}. Our study simulated a mathematical speckle potential that is empirically possible by utilizing laser beams  \cite{num_spec}. We studied circular random repulsive potential in the two-dimensional case. We also simulated a one-dimensional study on our system with linear random repulsive potential. We compared our results with a previous work \cite{2d_sahu} where square impurity was introduced.

\section{II. System \& Method of Calculations}
We have studied the droplet in 1D and 2D. The GP equations in 1D and 2D are different as the LHY corrections strongly depend on the dimensions of the system. 

\subsection*{Two dimensional system}
To create the quantum droplet,
we need to take into account the LHY correction term in addition to the mean-field terms.  We may write the GP equation for quantum droplet of Bose-Bose mixture
 with speckle potentials in two dimensions  \cite{2D_liquid, 2D_liquid4}

{\small
\begin{eqnarray}
   i\frac{\partial\psi_1 }{\partial t}=\left[-\frac{\nabla ^2}{2} + g(|\psi_1|^2 - |\psi_2|^2) + \frac{g^2}{4\pi} n \ln (n) + V(\vec{r}) \right ]\psi_1 \nonumber \\
      i\frac{\partial\psi_2 }{\partial t}=\left[-\frac{\nabla ^2}{2} + g(|\psi_2|^2 - |\psi_1|^2) + \frac{g^2}{4\pi} n \ln (n)+ V(\vec{r}) \right ]\psi_2. \\ \nonumber 
\end{eqnarray}
}
Where $n = |\psi_1|^2+|\psi_2|^2$ and $g$ denote the density of the droplet and the interaction strength between Bose atoms, respectively. The first term on the right side of the GP equation represents the kinetic energy; the second term denotes the mean-field interaction potential; the third term handles quantum fluctuation. We have considered a symmetric case with $\psi_1=\psi_2 \equiv \varphi/\sqrt{2}$. The reduced single GP   equation  is given by \cite{2D_liquid3}

\begin{eqnarray}
   i\frac{\partial\varphi }{\partial t}=\left[-\frac{\nabla ^2}{2}  + \frac{g^2}{4\pi} |\varphi|^2 \ln (|\varphi|^2) + V(\vec{r}) \right ]\varphi.
   \label{2Dsinglemode}
\end{eqnarray}

\begin{figure}
  \begin{center}
 \includegraphics[width=00.23\textwidth]{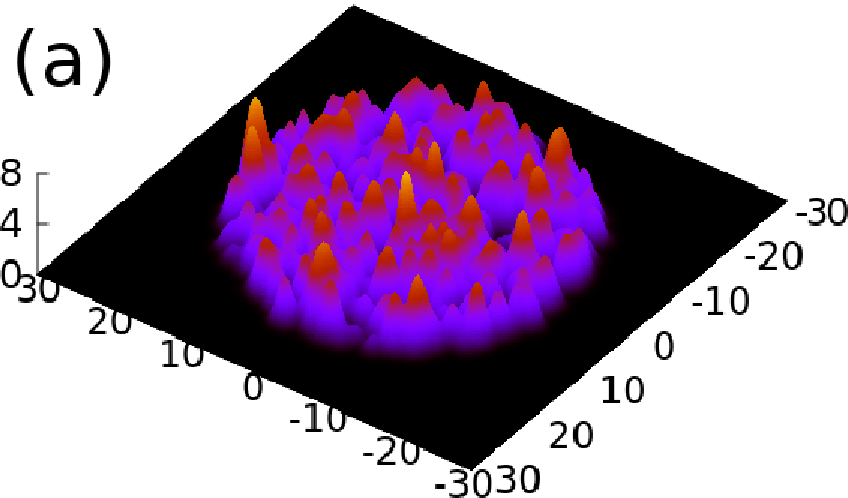}
 \includegraphics[width=00.23\textwidth]{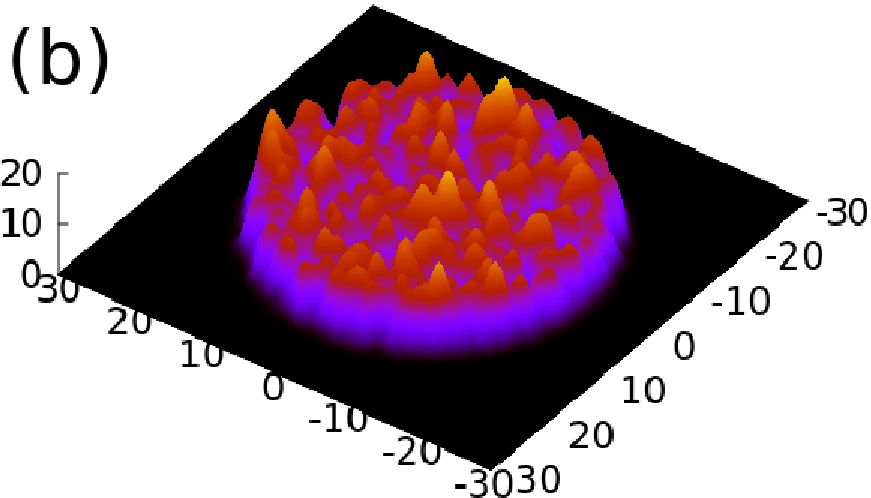} 
 \rule{0.4\textwidth}{0.4pt}
\includegraphics[width=00.23\textwidth]{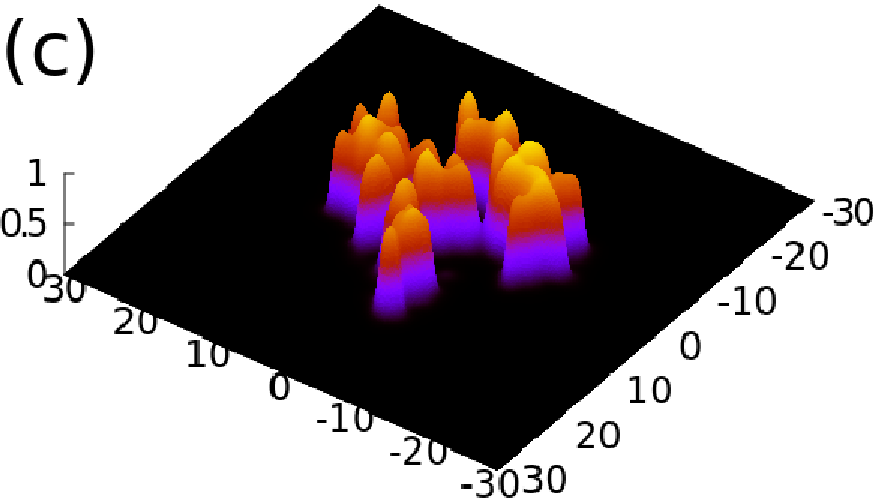}   
\includegraphics[width=00.23\textwidth]{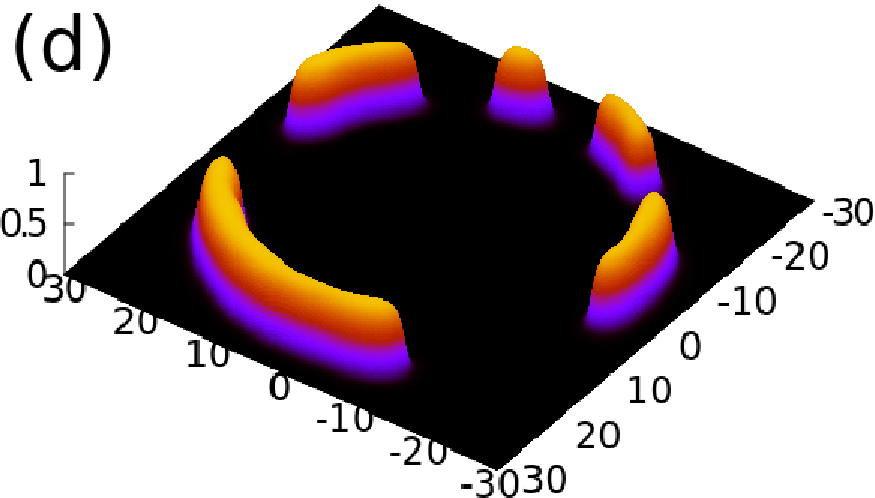}   

	  \caption{\textbf{(2D case):} (Top row) Circular impurity with different impurity densities (in $l_0^{-2}$ unit) $\rho  = 0.7956$ $(a)$, $3.1830$ $(b)$  ( equation (\ref{2d_dis})). (Bottom row ($c$ and $d$)). Droplet's density ($|\varphi|^2[l_0^{-2}]$) with the same impurity in the top row with $N =100$  and $g = 10$.  Droplet becomes porous (figure.1(c)) in nature for impurity density $\rho = 0.7956$. When $\rho$ increases further, the droplet (figure 1(d)) goes to the circumference of the impurity region (figure 1(b)).  }
    \label{fig:2Dpotl}
  \end{center}
\end{figure}

 Normalization of the wave function is given by

 \begin{eqnarray}
    \int |\varphi|^2 d^2\vec{r} = N.
 \end{eqnarray}

 \begin{figure}
  \begin{center}
    \includegraphics[height = 3.5 cm , width= 3.5cm]{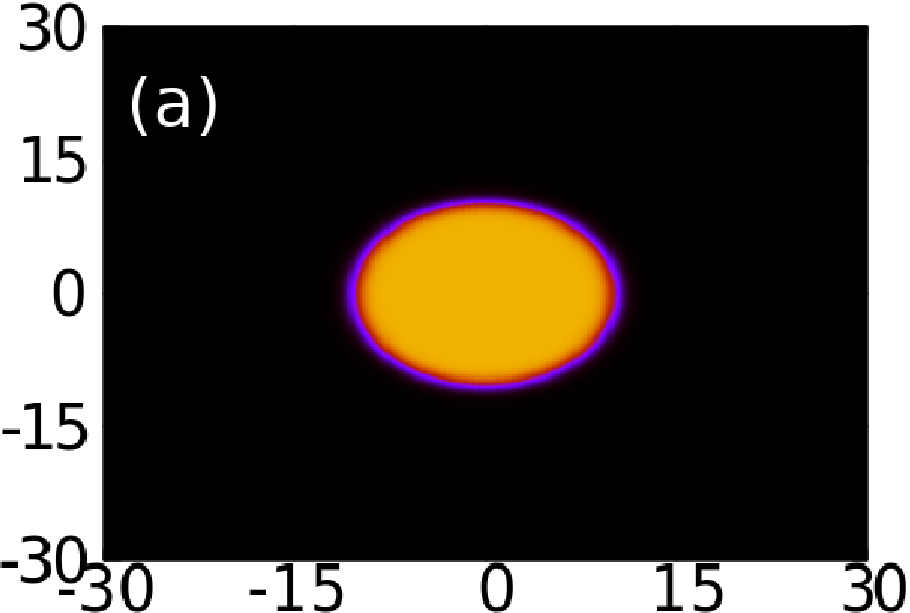}  
    \includegraphics[height=3.2cm]{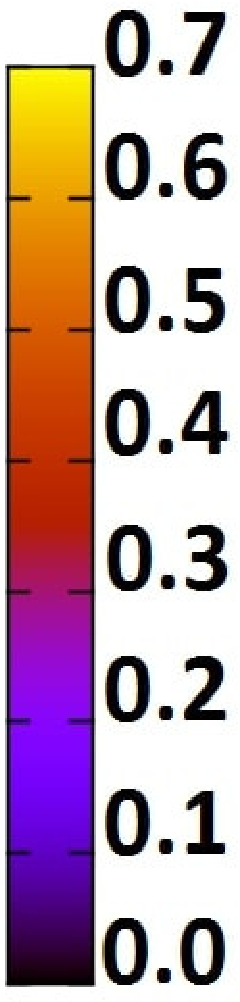} \\
    \includegraphics[height = 3 cm ,width=3cm]{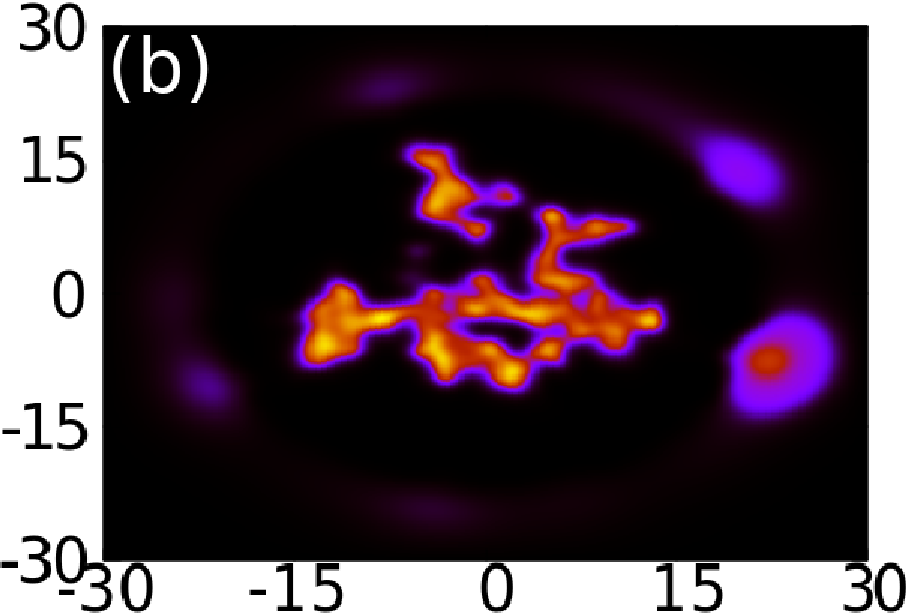} 
        \includegraphics[height = 3 cm ,width=3cm]{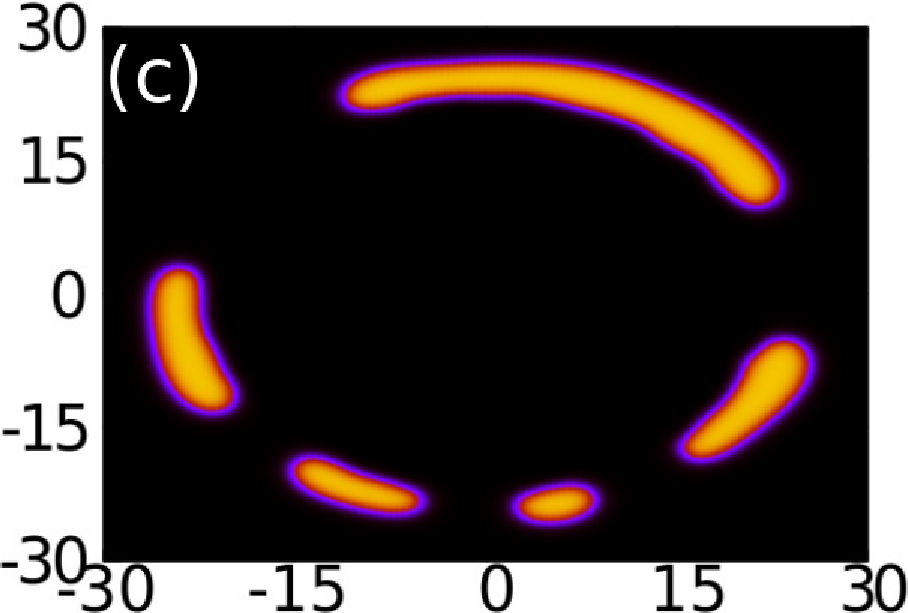}
            \includegraphics[height = 3 cm ,width=3cm]{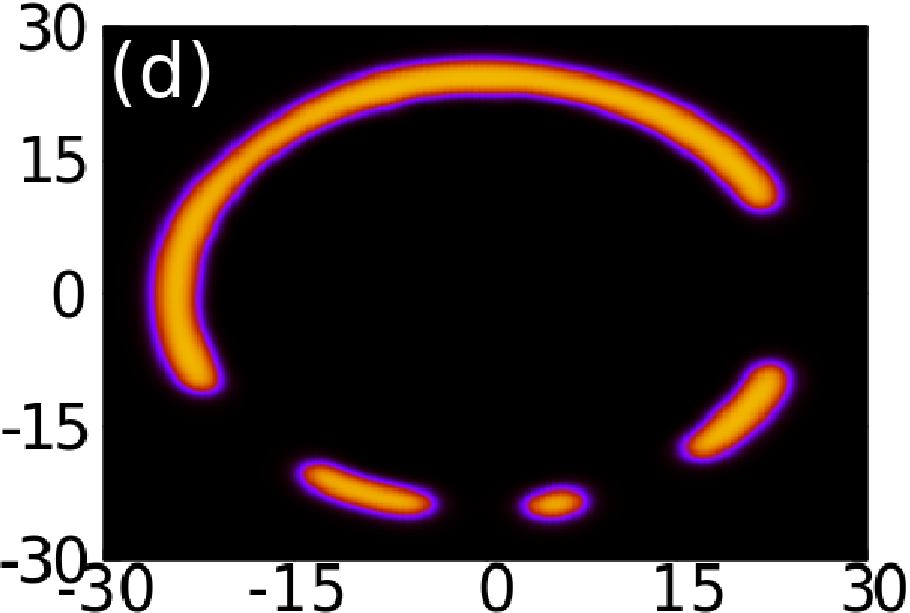}  
                \includegraphics[height = 3 cm ,width=3cm]{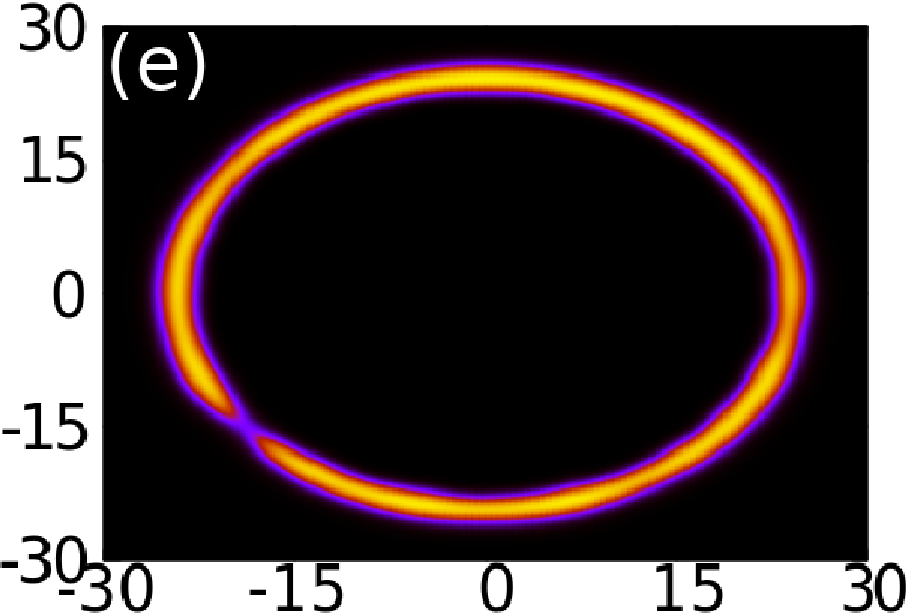} 
                    \includegraphics[height = 3 cm ,width=3cm]{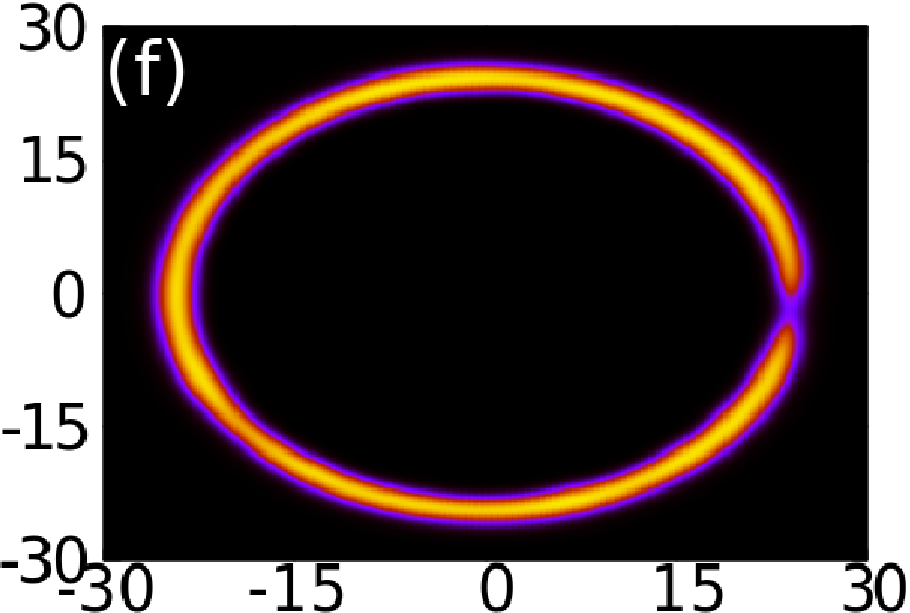} 
                        \includegraphics[height = 3 cm ,width=3cm]{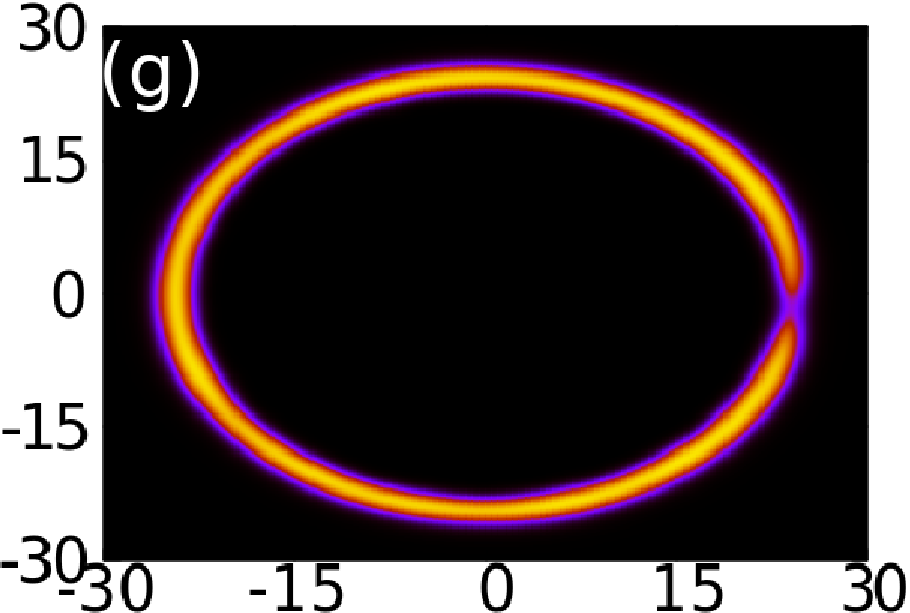} 
        
\caption{ \textbf{(2D case):} The droplet's density ($n[l_0^{-2}] = |\varphi|^2[l_0^{-2}]$): Impurity-free droplet is shown at the topmost figure $(a)$. Density of the droplet with different  impurity density ($\rho[l_0^{-2}]$) is shown by the figures $(b)$ to $(g)$. $\rho=$1.5915 $(b)$, 3.1830 $(c)$, 4.7745 $(d)$, 6.3660 $(e)$, 7.9575  $(f)$, and 9.549 $(g)$. The number of atoms is $N = 100$, and  $g = 10$ is the interaction strength.  It is studied that with a higher impurity density ($\rho$), the droplet goes to the circumference of the circular impurity and looks ring-shaped.}
    \label{fig:2Dg10den}
  \end{center}
\end{figure}

Where $N$ is the number of the condensed atom, the GP equation is expressed in scaled form \cite{dimension1,dimension2}, where the length scale is  $l_0$ (we have chosen $l_0=1.0\mu  $m \cite{dimension1} for numerical convenience), the unit of energy and time is $\hbar^2 /(m l_0^2$)  and $ml_0^2/\hbar$ \cite{dimension1}, respectively, where $m$ is the mass of condensed atom. Droplet generation may be described by two-body interaction, and our system's density is low enough to avoid three-body interaction. There have already been theoretical ideas for liquid BEC states in lower dimensions with only two-body interactions \cite{2D_liquid,2D_liquid1, 2D_liquid3,avra_PT}. We studied the Gaussian type \cite{Gaussian_Adhikari} of  speckle potential with random impurity points \cite{Gaussian2018, Gaussian2014}

 \begin{equation}
   V (\vec{r}) = V_0 \sum_{i = 1}^P  e^{(-(x-x_i)^2-(y-y_i)^2)/\chi ^2}. 
   \label{2d_dis}
 \end{equation}
 
 \begin{figure}[htbb]
  \begin{center}
	   \includegraphics[width=0.48\textwidth]{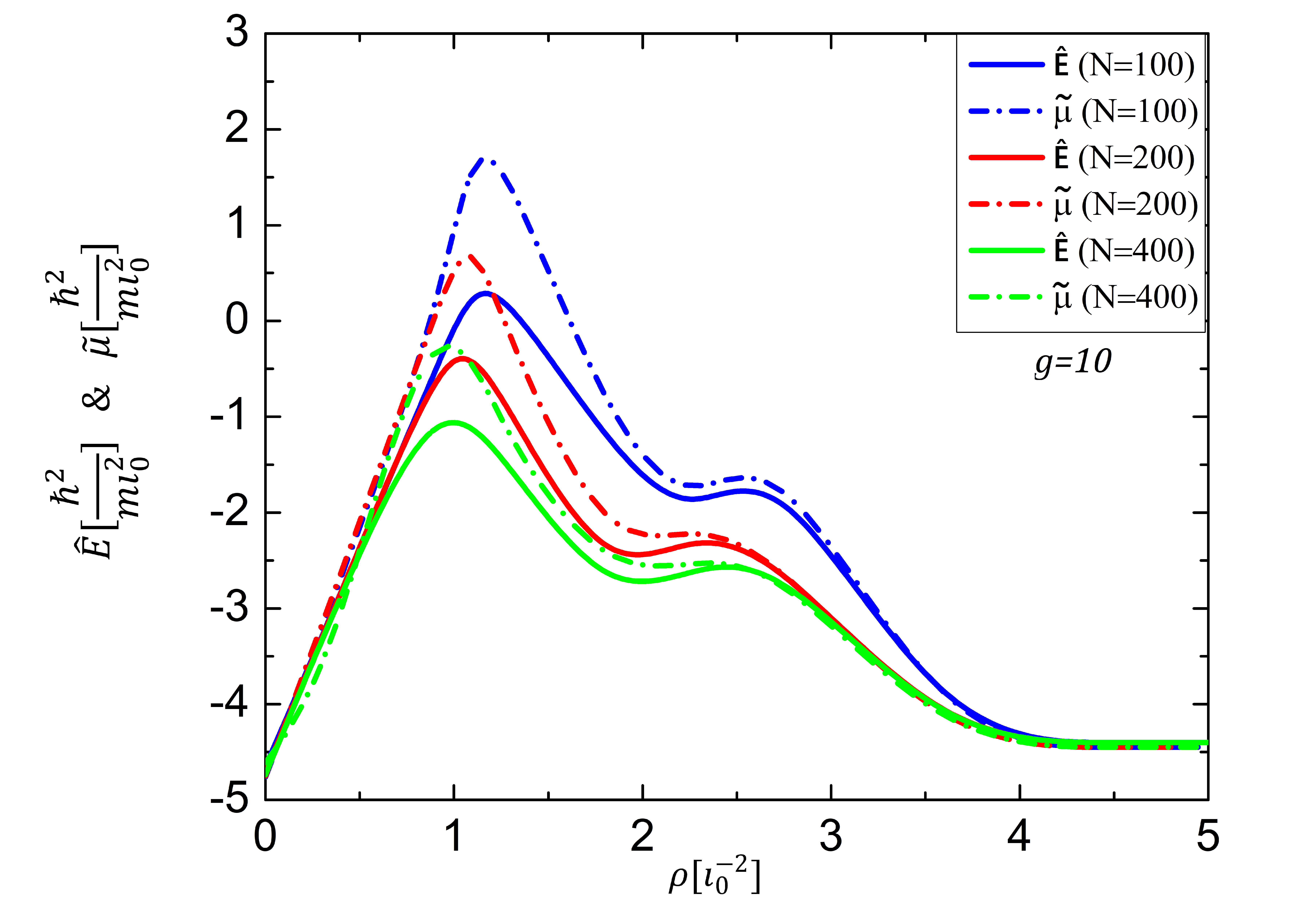}
	    \includegraphics[width=0.48\textwidth]{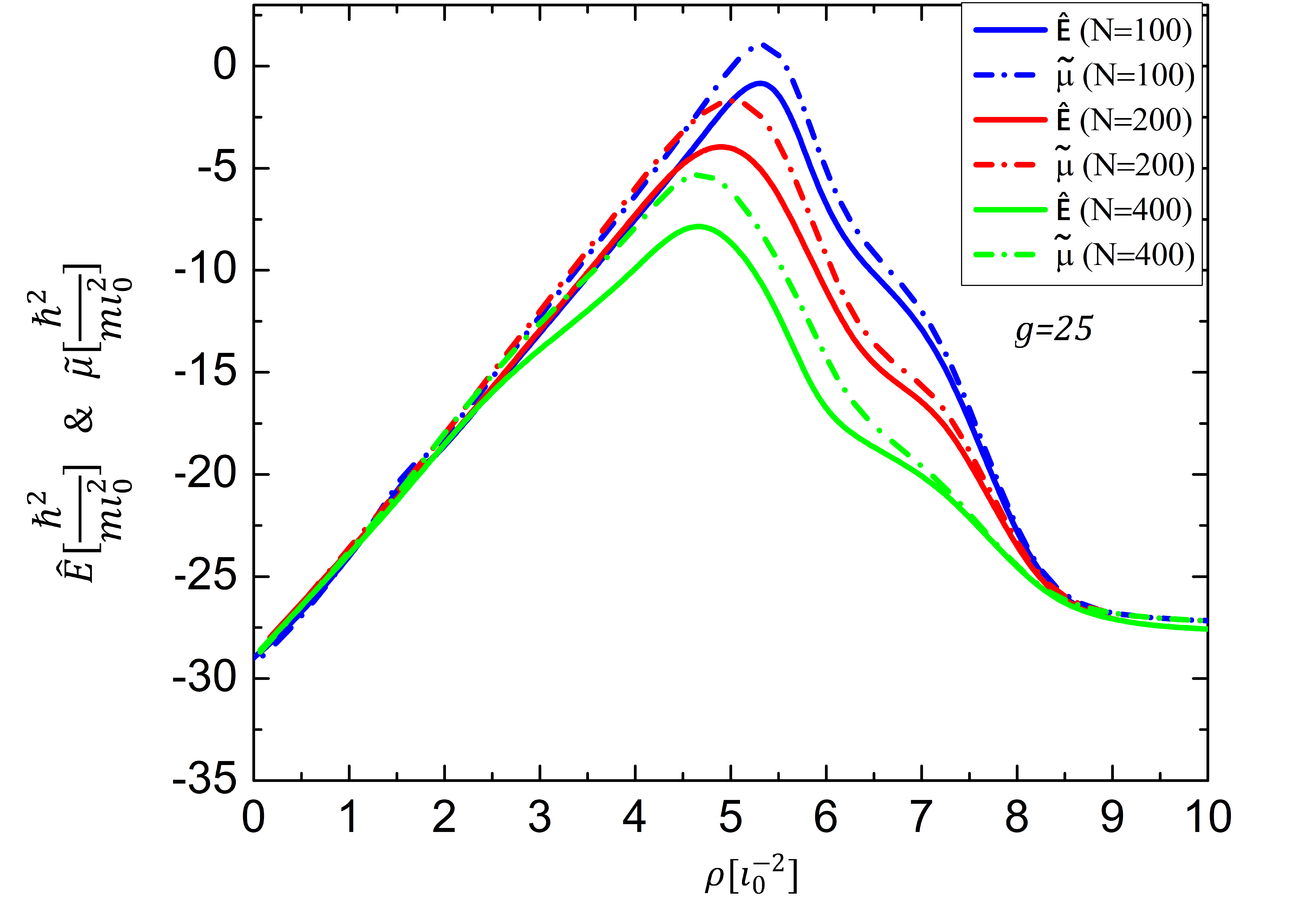}
\caption{ \textbf{(2D case):} The solid line is used to show the variation of energy per particle ($\hat{E} = \tilde{E}/N$), and the dashed line is used to show the variation of chemical potential ($\tilde{\mu}$) for various values of impurity density ($\rho$), $N = 100,\,200,\,400$ and $g = 10$  and $g =25$.  Both $\hat{E}$ and $\tilde{\mu}$ increase with impurity density ($\rho$) and reach a maximum value. The density corresponding to this maximum energy is porous, with the maximum surface area exposed to the impurity. If $\rho$ increases further, porous droplets start to move out of the impurity region. Energy starts to decrease and reach a saturated value. Still, this final energy ($\hat{E}$ at $\rho$ = 5 for $g $ = 10) is higher than the initial value ($\hat{E}$ at $\rho$ = 0 for $g $ = 10) as finally the droplet's surface area and surface energy are larger than the initial. Higher impurity density ($\rho$) is required to get maximum energy and chemical potential when interaction strength ($g$) increases. }
    \label{fig:2Denergy}
  \end{center}
\end{figure}

 The strength and the characteristic length of speckle potential ($V$) are denoted by $V_0$ and $\chi$, respectively, and  $V_0$ and $\chi$ are taken unity in our calculations. The number of impurity points is $P$ within the considered circular area ($\pi r^2$) of speckle potential. Impurity density of speckle potential is $\rho = P/(\pi r^2)$  (figure (\ref{fig:2Dpotl}) (Top row)). Here, the considered impurity is circular with radius $r =20$, and the boundary ($60 \times 60$) is large enough to affect the system (figure (\ref{fig:2Dpotl})). Simulations are performed with the droplet at the central zone of the impurity. Experimentally, the random impurity points \{($x_i,y_i$)\}  can be created using  lasers \cite{speckle1}. 
 The GP equation (\ref{2Dsinglemode}) is solved using the Split Step Crank-Nicolson (CN) method  \cite{fortran, c sadhan,dm_coll}. The GP equation is solved for smaller values  $V_0$ with unperturbed wave function as an initial guess, and then $V_0$ is increased and solved the GP equation with wave function obtained in the last step as the initial guess. The droplet's calculated density is shown in figure (\ref{fig:2Dg10den}) with interaction strength $g = 10$.

We  calculated the energy and chemical potential after finding the wave function using the formulas \cite{2D_liquid3,2D_liquid4}

{\small
\begin{eqnarray}
E &=& \int \Big ( \frac{1}{2} \left(|\nabla\varphi|^2 \right)+ \frac{g^2} {8\pi} |\varphi|^4 \ln \Big (\frac{|\varphi|^2}{{\sqrt e}} \Big ) + V(\vec{r})|\varphi|^2  \Big )d^2\vec{r} \nonumber \\
\mu &=& \frac{1}{N}\int \Big ( \frac{1}{2} \left(|\nabla\varphi|^2 \right) + \frac{g^2} {4\pi} |\varphi|^4  \ln (|\varphi|^2)  + V(\vec{r})|\varphi|^2  \Big )d^2\vec{r}. \nonumber \\
\end{eqnarray}
}
Where $e$ is the base of the natural logarithms, the disordered average ($\tilde{E}$ and $\tilde{\mu}$) is obtained by taking different sets of impurity point configurations for a given density of impurity  ($\rho$). The average energy and chemical potential of the system with disordered potential is given by

\begin{eqnarray}
	\tilde{E} = \frac{1}{N_P}\sum^{N_P}_{i=1}   E_i \nonumber \\
	\tilde{\mu} = \frac{1}{N_P}\sum^{N_P}_{i=1}   \mu_i.
	\label{average}
\end{eqnarray}

Where $N_P$ is the total number of impurity configurations with a given impurity density ($\rho$), and it is sufficient to take $N_P = 20$ (two simulations with all the same parameters give slightly different energy and chemical potential values due to the randomness of the potential. To overcome this, we have taken average values of sufficient configurations). The energy and chemical potential for a given impurity density are  $E_i$ and $\mu_i$, respectively. 

\subsection*{One dimensional system}

In one-dimensional system, the reduced single GP equation with LHY interaction for quantum droplet of Bose-Bose mixture is given by \cite{2D_liquid}
\begin{eqnarray}
   i\frac{\partial\psi }{\partial t}=-\frac{1}{2}\nabla ^2\psi + \delta g|\psi|^2\psi - \frac{\sqrt{2}}{\pi} g^{3/2}|\psi|\psi + V(x)\psi. 
   \label{1dsinglemode}
\end{eqnarray}

The normalization condition  is given by \\

\begin{eqnarray}
  \int |\psi|^2 dx = N.
\end{eqnarray}

 Where $N$ is the total number of condensed atoms, the GP equation is written using $\delta g$ for calculation convenience. It is defined as $\delta g = g_{12} + \sqrt{g_{11}g_{22}}$, where $g_{12}$ is the interaction between two different species and $g_{11}, g_{22}$ is the interaction between the same species of atoms and in our calculation, we have used numerical value of $\delta g=g$. The first term on the right side of the GP equation represents kinetic energy; the second term is the mean-field interaction; the third term is the quantum fluctuation (LHY) term in 1D, and the last term is for the random speckle potential. The speckle potential  has been taken  as Gaussian function type and is given by
 
 
 \begin{equation}
   V (x) = V_0 \sum_{i = 1}^P  e^{-(x-x_i)^2/\chi^2}.
 \end{equation}
 
  \begin{figure}[htbb]
  \begin{center}
           \includegraphics[width=0.51\textwidth]{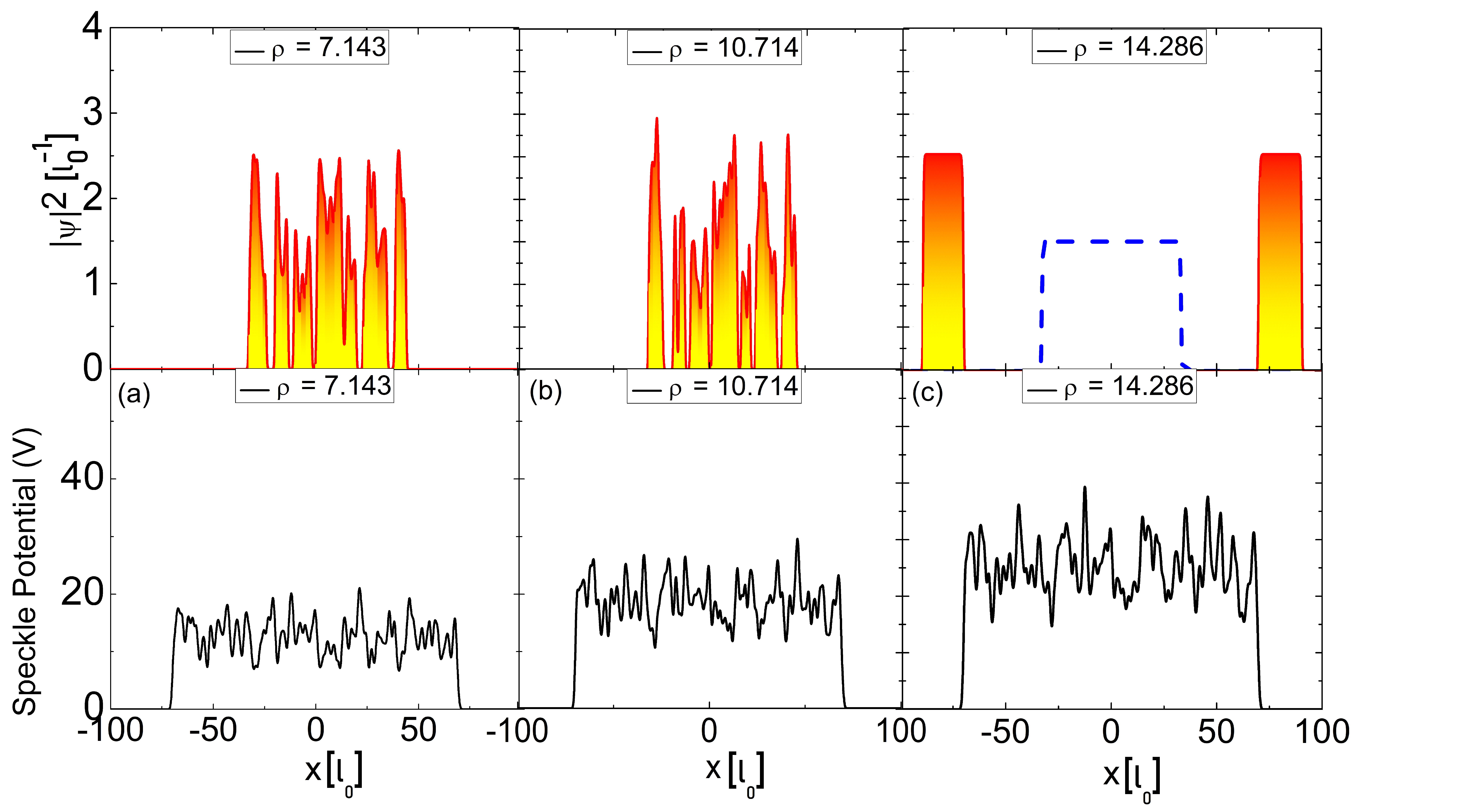}
          
                      \includegraphics[width=0.51\textwidth]{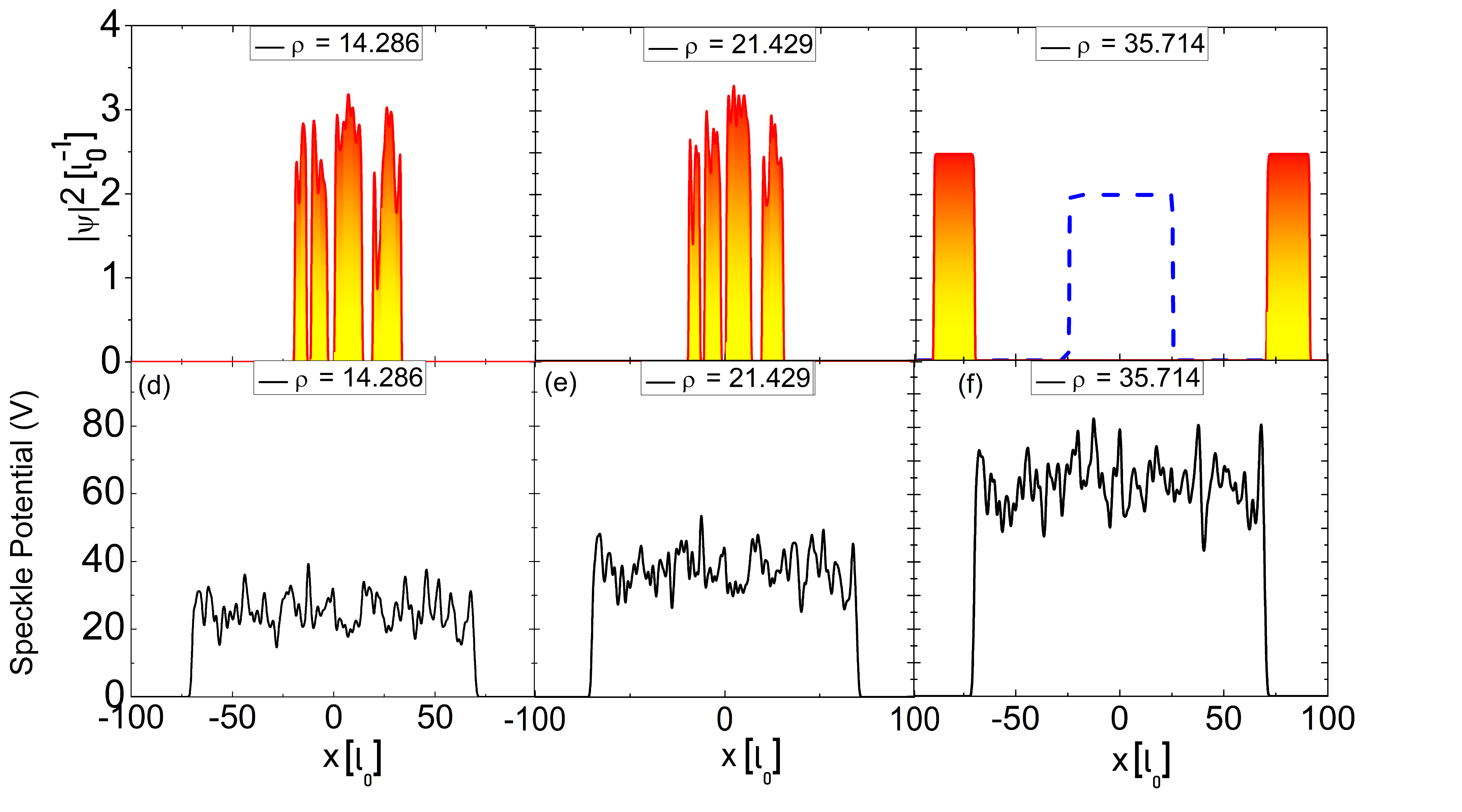} 
\caption{ \textbf{(1D case):} The droplet's density ($|\psi|^2$) in presence of random speckle potential for various  $\rho[l_0^{-1}]$, $g = 10$ (Top panel) and $g = 20$ (Bottom panel) for $N = 100$ particles. We have chosen $\delta g = g$ in each case. The blue dashed line represents the density of the droplet without impurity potential. The corresponding impurity potential is shown below the density.   We studied that the droplet becomes porous for smaller impurity density ($\rho$) and moves out of the impurity region for higher $\rho$. Higher impurity density ($\rho$) is required to separate droplet when interaction strength ($g$) increases.}   
   \label{fig:1Dg10den}
     \end{center}
\end{figure}

The speckle potential's strength and the characteristic length are  $V_0$ and $\chi$, respectively, and  $V_0$ and $\chi$ are taken unity in our calculations. We have used random impurity points  \{$x_i$\}. The number of impurity points is $P$  inside the length $L=140$, and mesh size ($200$, figure \ref{fig:1Dg10den}) is large enough to affect the system's energy and chemical potential. The density of the impurity points is $\rho = \frac{P}{L}$. Simulations are performed with the droplet at the central zone of impurity. The Split Step Crank-Nicolson (CN) method is used to solve the GP equation \cite{fortran, c sadhan,dm_coll}. Here, we are applying the same steps as those in 2D to calculate the density, energy $(E)$, and chemical potential $(\mu)$  of the system \cite{2D_liquid}. 

\begin{small}
\begin{eqnarray}
\nonumber E &=&  \int \left(\frac{1}{2}(|\nabla\psi|^2 + \delta g|\psi|^4 - \frac{4\sqrt{2}}{3\pi} g^{3/2}|\psi|^3 + V(x)|\psi|^2 \right) dx  \\
\nonumber \mu &=& \frac{1}{N} \int \left(\frac{1}{2}(|\nabla\psi|^2 + \delta g|\psi|^4 - \frac{\sqrt{2}}{\pi} g^{3/2}|\psi|^3 + V(x)|\psi|^2 \right) dx. \\ 
\end{eqnarray}
\end{small}

We have calculated the disorder potential average of energy and chemical potential using equation (\ref{average}), as discussed earlier.

\begin{figure}[htbb]
  \begin{center}
           \includegraphics[height=0.355\textwidth]{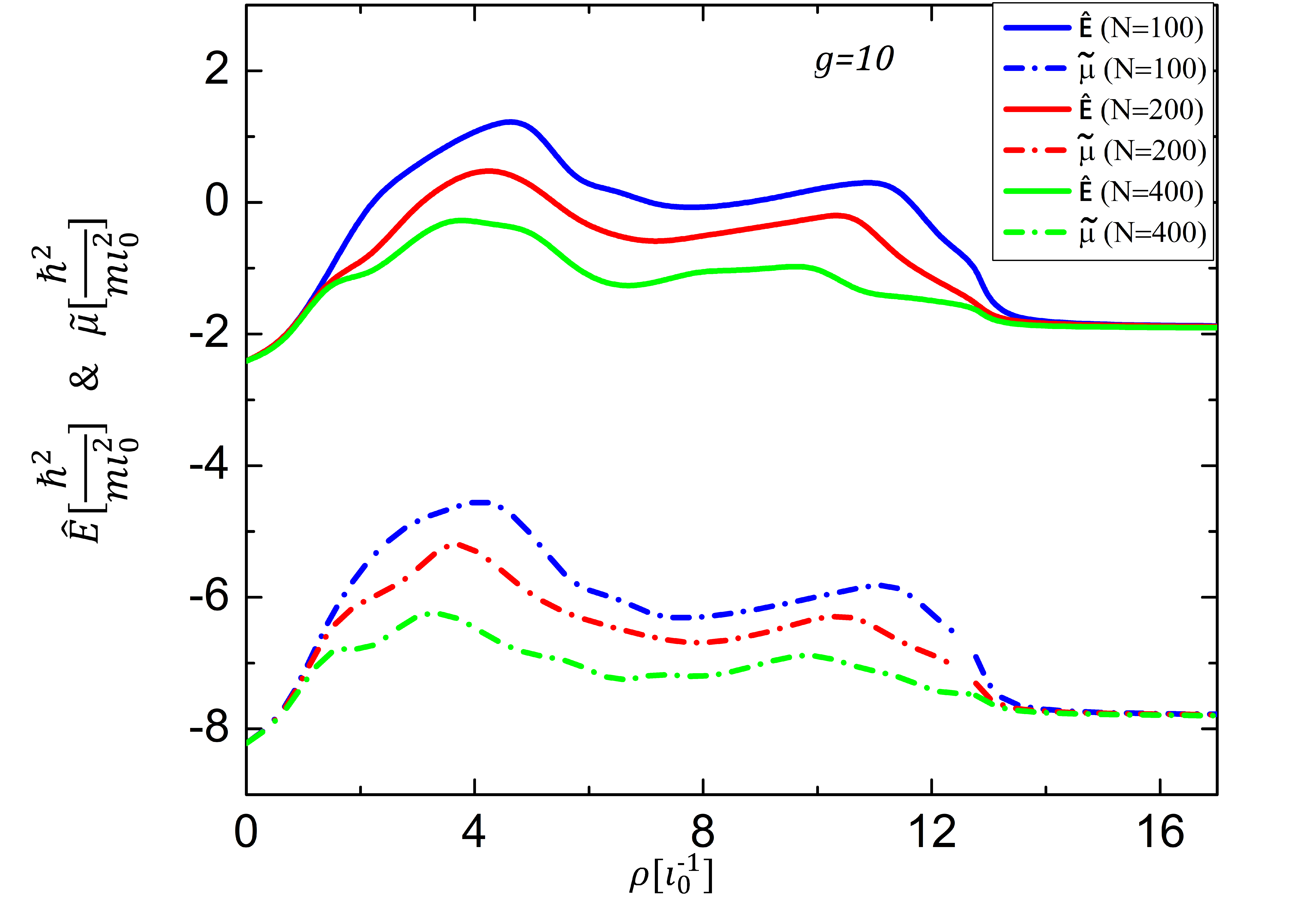} 
           \includegraphics[height= 0.354\textwidth]{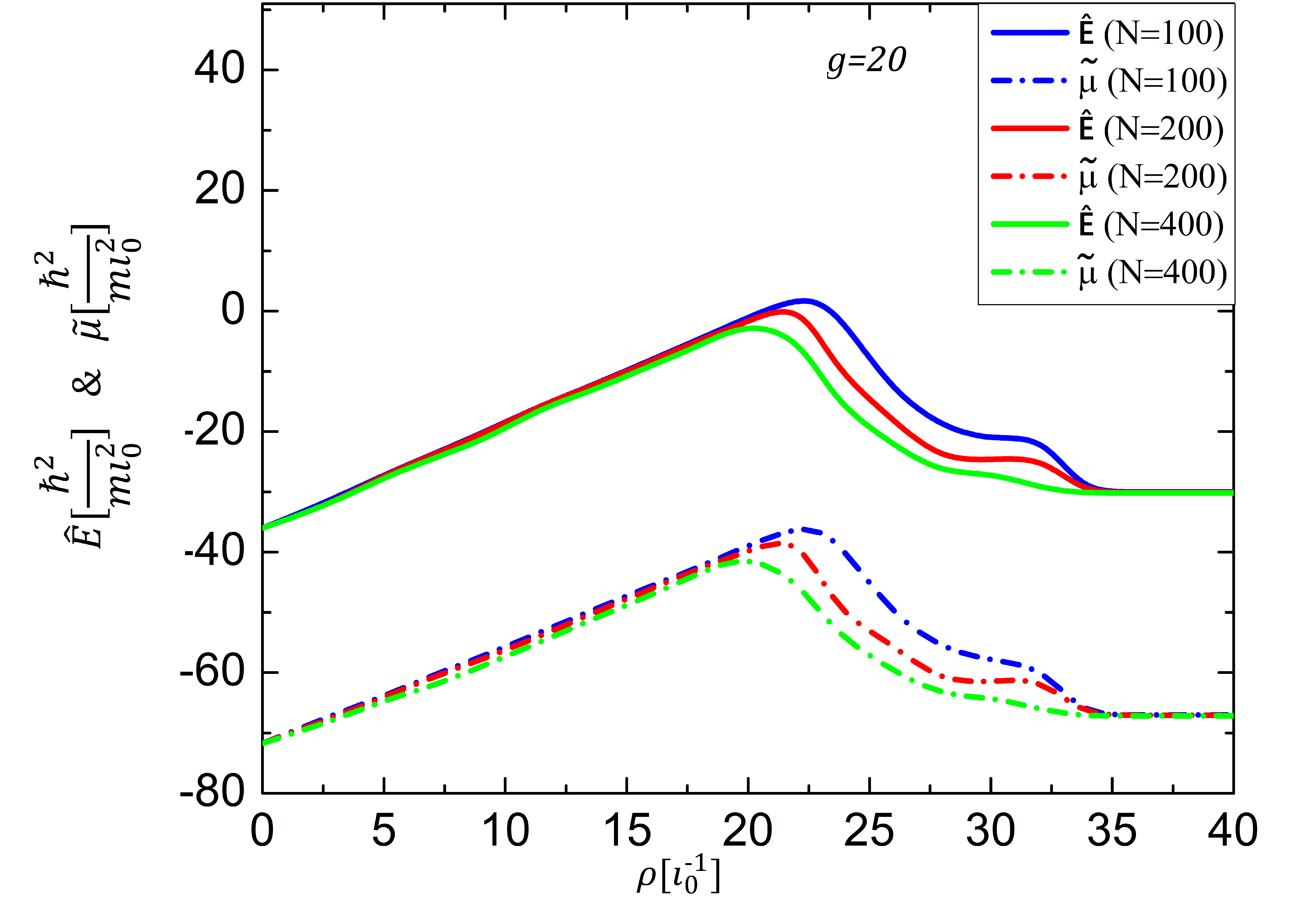}   
\caption{ \textbf{(1D case):} The solid line is used to show the variation of energy per particle ($\hat{E} = \tilde{E}/N$), and the dashed line is used to show the variation of chemical potential ($\tilde{\mu}$) for various values of impurity density of speckle potential ($\rho$), $g = 10$ and $g = 20$ for $N =100,200 $ and  $ 400$.  Both $\hat{E}$ and $\tilde{\mu}$ increases with impurity density $\rho$ and reach a maximum value.  If $\rho$ increases further, porous droplets start to move out of the impurity region. Energy starts to decrease and reach a saturated value. Still, this final energy ($\hat{E}$ at $\rho$ = 16 for $g $ = 10) is higher than the initial value  ($\hat{E}$  at $\rho$ = 0 for $g $ = 10) as finally droplet's surface area and surface energy is larger than the initial. Higher impurity density ($\rho$) is required to get maximum energy and chemical potential when interaction strength ($g$) increases. }
    \label{fig:1Denergy}
  \end{center}
\end{figure}

\section{III. Results and discussion}

The qualitative nature of the result remains the same for one- and two-dimensional systems.
We studied the density of the condensate for two dimensions by solving equation (\ref{2Dsinglemode}) with the strength of interactions $g =10,\, 25$ and for one dimension by solving equation (\ref{1dsinglemode}) with $g =10,\, 20$ for various impurity densities. Density profiles for 2D are illustrated in figures (\ref{fig:2Dpotl}) and (\ref{fig:2Dg10den}) and for 1D in figure (\ref{fig:1Dg10den}). In both cases, we obtained a sharp droplet without the impurity potential (for 2D, it is shown in the topmost figure of figure (\ref{fig:2Dg10den}) and for 1D, it is shown by the blue dashed line in figure (\ref{fig:1Dg10den})). Then, we introduce the impurity speckle potential and porous droplet forms, and when the impurity density is increased, the droplet begins to split up. The split in the density is more prominent when the impurity density of speckle potential is higher. The droplet's density demonstrates that the BEC is still in the liquid phase inside the disorder zone.

The condensate travels out of the potential region when the impurity density is increased further. The crucial aspect is that the BEC remains in the liquid phase in the presence of a high impurity potential; otherwise, in a repulsive external potential, the atoms will leave the condensate. The condensate behaves similarly to a traditional liquid droplet. It can also be concluded that the system with higher interaction strength ($g=20$) can withstand a higher impurity potential than the one with $g=10$. For the 2D system, as we considered a circularly arranged random potential, the resultant density profiles look like rings. \\


We displayed the system's energy and chemical potential  ($g=10,\,25$) for various values of impurity density in figure(\ref{fig:2Denergy}) for the 2D system. Figure (\ref{fig:1Denergy}) represents the energy and chemical potential for the 1D system. Energy and chemical potential grow with the increase in speckle density, reach a maximum, then decrease and become constant. Speckle potential is repulsive, so the increment of energy and chemical potential with impurity density is well expected. After a certain impurity density, the energy and chemical potential begin to decrease. The droplet gets fragmented into small droplets with higher impurity density, and some droplets travel out of the impurity zone; this is the reason for the decrease in energy. 

All droplets move out of the potential zone after particular values of impurity density. In this case, the disorder potential does not affect the condensate.  At this point, energy and chemical potential values are somewhat greater than those of the system without impurities. This is due to the droplet being separated from the potential zone and forming two droplets (1D) and ring-shaped (2D) droplets with a larger total surface area than the initial single droplet.
Without speckle potential, the energy per particle is independent of condensate size; however, in speckle potential, the energy per particle declines with condensate size. 

In conclusion, we have studied the effects of the circular and linear random speckle potential on the quantum droplet in 2D and 1D, respectively. As the LHY term is dimensionally dependent, we have studied 1D and 2D separately. We found that in 2D, the final density (figure \ref{fig:2Dg10den}(g)) is ring-shaped, unlike the density due to the square impurity in reference \cite{2d_sahu}. In the 1D study, random impurity potentials divide the droplet into two symmetric droplets as impurity potentials are generated like figures \ref{fig:1Dg10den}(c) and \ref{fig:1Dg10den}(f). Overall, the qualitative nature of the chemical potential and energy are equivalent to reference \cite{2d_sahu}, but the variation with impurity density is different, as here in 2D circular impurity is considered. 1D energy and chemical potential variations are different due to different LHY and disorder potential.\

\textbf{Author contributions} All the authors contributed significantly to the study. Both the authors, Avra Banerjee and Saswata Sahu, had the lead role in the study. Dwipesh Majumder supervised the whole work.\\

\textbf{Funding }This research was supported by Indian Institute of Engineering Science and Technology, Shibpur.

\textbf{Data Availability Statement} No data associated in the manuscript.\\

\textbf{Declarations}\\

\textbf{Conflict of interest} The authors declare that they have no Conflict of interest in the publication of this paper.


\textbf{Ethical approval} All authors approved the manuscript and possible publication.


\begin{thebibliography}{50}


\bibitem{bec1}A. J. Leggett, Rev. Mod. Phys. {\bf 47}, 331 (1975).
\bibitem{bec2} F. Dalfovo, S. Giorgini, L. P. Pitaevskii, and S. Stringari, Rev. Mod. Phys. {\bf 71}, 463 (1999).
\bibitem{Petrov2015} D. S. Petrov, Phys. Rev. Lett. {\bf 115}, 155302 (2015).
\bibitem{LHY1} T. D. Lee and C. N. Yang, Phys. Rev. {\bf 105}, 1119 (1957).
\bibitem{LHY2} T. D. Lee, K. Huang and C. N. Yang, Phys. Rev. {\bf 106}, 1135 (1957).
\bibitem{LHY_3} A. Banerjee and D. Majumder, Annals of Physics {\bf 470}, 169810 (2024).
\bibitem{Trarruell2018} C. R. Cabrera, L. Tanzi, J. Sanz, B. Naylor, P. Thomas, P. Cheiney, L. Tarruell, Science {\bf 359}, 301 (2018).
\bibitem{drop_exp2} G. Semeghini, G. Ferioli, L. Masi, C. Mazzinghi, L. Wolswijk, F. Minardi, M. Modugno, G. Modugno, M. Inguscio, M. Fattori,  	Phys. Rev. Lett. {\bf 120}, 235301 (2018).
\bibitem{Trarruell2018PRL} P. Cheiney, C. R. Cabrera, J. Sanz, B. Naylor, L. Tanzi, L. Tarruell, Phys. Rev. Lett. {\bf 120}, 135301 (2018). 
\bibitem{homo} G. Ferioli, G. Semeghini, L. Masi, G. Giusti, G. Modugno, M. Inguscio, A. Gallem\'{i}, A. Recati and M. Fattor, Phys. Rev. Lett. {\bf 122}, 090401 (2019).
\bibitem{Rb_Na} C. D'Errico, A. Burchianti, M. Prevedelli, L. Salasnich, F. Ancilotto, M. Modugno, F. Minardi and C. Fort, Phys. Rev. Research {\bf 1}, 033155 (2019).
\bibitem{Itali2020} A. Burchianti, C. D. Errico, M. Prevedelli, F. Ancilotto, M. Modugno, L. Salasnich, F. Minardi and C. Fort, Condens. Matter {\bf 5}, 21 (2020).
\bibitem{dipolar_droplets}Observation of dipolar Q.D.: H. Kadau, M. Schmitt, M. Wenzel, C. Wink, T. Maier, I. F. Barbut, and T. Pfau, Nat. Phys. {\bf 530}, 194 (2016); 
I. F. Barbut, H. Kadau, M. Schmitt, M. Wenzel, and T. Pfau, Phys. Rev. Lett. {\bf 116}, 215301 (2016); I. F. Barbut, M. Schmitt, M. Wenzel, H. Kadau, and T. Pfau, J. Phys. B {\bf 49}, 214004 (2016).
\bibitem{3body2} S. K. Adhikari, Phys. Rev. A {\bf 95}, 023606 (2017).
\bibitem{bog_1_lhy} V. N. Popov, Teor. Mat. Fiz. {\bf 11}, 565 (1972). 
\bibitem{bog_2_lhy} E. H. Lieb and W. Liniger, Phys. Rev. {\bf 130}, 1605 (1963).

\bibitem{2D_liquid} D. S. Petrov and G. E. Astrakharchik, Phys. Rev. Lett. \textbf{117}, 100401 (2016).
\bibitem{2D_liquid1} G. E. Astrakharchik, B. A. Malomed, Phys. Rev. A \textbf{98}, 013631 (2018).
\bibitem{2D_liquid2} P. Zin, M. Pylak, T. Wasak, M. Gajda, Z. Idziaszek, Phys. Rev. A \textbf{98}, 051603(R) (2018).
\bibitem{1D2D2018} Dimensional crossover for the beyond-mean-field correction in Bose gases, T. Ilg, J. Kumlin, L. Santos, D. S. Petrov and H. P. Büchler, Phys. Rev. A {\bf 98}, 051604(R) (2018).  
\bibitem{2D_liquid3} Y. Li, Z. Chen, Z. Luo, C. Huang, H. Tan, W. Pang, B. A. Malomed, Phys. Rev. A {\bf 98}, 063602 (2018).
\bibitem{2D_liquid4}Y. Li, Z. Luo, Y. Liu, Z. Chen, C. Huang, S. Fu, H. Tan, and B. A. Malomed, New J. Phys. 19, 113043 (2017).
\bibitem{Gajda19} D. Rakshit, T. Karpiuk, P. Zin, M. Brewczyk, M. Lewenstein, M. Gajda, New J. Phys. \textbf{21}, 073027 (2019).

\bibitem{low_Gajda2023} P. Zin, M. Pylak and M. Gajda, Phys. Rev. A \textbf{106}, 013320 (2022).

\bibitem{leonard} L. Fallani, C. Fort, and M. Inguscio, Adv. At. Mol. Opt. Phys. \textbf{56}, 119 (2008).

\bibitem{speckle1} J. E. Lye, L. Fallani, M. Modugno, D. S. Wiersma, C. Fort, and M. Inguscio, Phys. Rev. Lett. {\bf 95}, 070401 (2005). 
\bibitem{speckle2} R. C. Kuhn, O. Sigwarth, C. Miniatura, D. Delande and C. A. Muller, New Journal of Phys. {\bf 9}, 161 (2007). 
\bibitem{speckle3} B. Abdullaev and A. Pelster, Eur. Phys. J. D {\bf 66}, 314 (2012); A. Boudjemaa, Phys. Rev. A {\bf 91}, 053619 (2015).
\bibitem{speckle4} S. Pilati and P. Pieri, scientific reports, {\bf 9}, 5613 (2019).
\bibitem{speckle5} Y. P. Chen, J. Hitchcock, D. Dries, M. Junker, C. Welford, S. E. Pollack, T. A. Corcovilos, R. G. Hulet, Physica D {\bf 238}, 1321 (2009).



\bibitem{Gaussian1} M. Kobayashi and M. Tsubota, Phys. Rev. B {\bf 66}, 174516 (2002); G. M. Falco, A. Pelster and R. Graham, Phys. Rev. A {\bf 76}, 013624 (2007); C. Krumnow and A. Pelster, Phys. Rev. A {\bf 84}, 021608 (2011).
\bibitem{Gaussian_Adhikari} Y. Cheng, S. K. Adhikari, Phys. Rev. A {\bf 82} 013631 (2010).
\bibitem{GP_random2} E. Akkermans, S. Ghosh and Z. Musslimani, J. Phys. B {\bf 41}, 045302 (2008). 
\bibitem{Gaussian2014} M. Sajid, I. Ashraf, Laser Phys. {\bf 24}, 115501 (2014). 
\bibitem{Gaussian2018} Sh. Mardonov, V. V. Konotop, B. A. Malomed, M. Modugno, and E. Ya. Sherman, Phys. Rev. A {\bf 98}, 023604 (2018).
\bibitem{Lorentzian} B. Nikolic, A. Balaz and A. Pelster, Phys. Rev. A {\bf 88}, 013624 (2013).

\bibitem{Anderson2007} L. Sanchez-Palencia, D. Clement, P. Lugan, P. Bouyer, G. V. Shlyapnikov, and A. Aspect, Phys. Rev. Lett. {\bf 98}, 210401 (2007).
\bibitem{Anderson2010} G. Modugno, Rep. Prog. Phys. {\bf 73}, 102401 (2010). 
\bibitem{Anderson2008} G. Roati, C. D'Errico, L. Fallani, M. Fattori, C. Fort, M. Zaccanti, G. Modugno, M. Modugno, M. Inguscio, Nature {\bf 453}, 895 (2008). 
\bibitem{Anderson2021} M. C. P. dos Santos and W. B. Cardoso, Phys. Rev. E {\bf 103}, 052210 (2021).

\bibitem{SIT3_2004} E. Altman, Y. Kafri, A. Polkovnikov, and G. Refael, Phys. Rev. Lett. {\bf 93}, 150402 (2004). 
\bibitem{SIT1_2014} C. D'Errico, E. Lucioni, L. Tanzi, L. Gori, G. Roux, I. P. McCulloch, T. Giamarchi, M. Inguscio, and G. Modugno, Phys. Rev. Lett. {\bf 113}, 095301 (2014). 
\bibitem{SIT2_2013} L. Tanzi, E. Lucioni, S. Chaudhuri, L. Gori, A. Kumar, C. D’Errico, M. Inguscio and G. Modugno, Phys. Rev. Lett. {\bf 111}, 115301 (2013). 
\bibitem{BBM-2020} A. Boudjemaa and K. Abbas, Phys. Rev. A {\bf 102}, 023325 (2020).
\bibitem{2d_sahu} S. Sahu, D. Majumder, Eur. Phys. J. Plus {\bf 137}, 1020 (2022).
\bibitem{boundImp1} P. Horak, J. Y. Courtois and G. Grynberg,  Phys. Rev. A {\bf 58}, 3953 (1998). 
\bibitem{PIM_2010_Pilati} S. Pilati, S. Giorgini, M. Modugno and N. Prokofev, New Journal of Physics {\bf 12}, 073003  (2010) (28pp). 

\bibitem{diff_mc} G. E. Astrakharchik, J. Boronat, J. Casulleras, and S. Giorgini, Phys. Rev. A. {\bf 66}, 023603v (2002).
\bibitem{quantum_mc} N. Laflorencie, Euro. Phys. Lett. {\bf 99}, 66001 (2012).
\bibitem{GP_random1} B. Min, T. Li, M. Rosenkranz and W. Bao, Phys. Rev. A {\bf 86}, 053612 (2012); 
 X. Antoine, W. Bao, C. Besse, Computer Physics Communication {\bf 184}, 2621 (2013). 
\bibitem{GP_random3} R. Acosta-Diaz, G. Krein, A. Saldivar, N. F. Svaiter and C. A. D. Zarro, J. Phys. A: Math. Thro. {\bf 52}, 445401 (2019). 

\bibitem{avra_PT} A. Banerjee and D. Majumder, Phys. Scr. {\bf 99}, 085402 (2024).
\bibitem{analytical_ran1} P. Lugan, D. Clement, P. Bouyer, A. Aspect, and L. Sanchez-Palencia, Phys. Rev. Lett. {\bf 99}, 180402 (2007). 

\bibitem{MFT1}  V. I. Yukalov and R. Graham, Phys. Rev. A {\bf 75}, 023619 (2007). 


\bibitem{num_spec} Y. Su, Q. Zhang, and Z. Gao, Opt. Express {\bf 25}, 30259 (2017).

\bibitem{dimension1} S. Gautam, S. K. Adhikari, J. Phys. B {\bf 52}, 055302 (2019).
 \bibitem{dimension2} S. Gautam, S. K. Adhikari, Annals of Phys. {\bf 409}, 167917 (2019).
\bibitem{dm_coll}  A. Banerjee, D. Majumder, J. Low Temp. Phys. {\bf 215}, 64 (2024).
\bibitem{fortran} P. Muruganandam, S. K. Adhikari, Comput. Phys. Commun. {\bf 180}, 1888 (2009).
\bibitem{c sadhan} D. Vudragovi\'{c}, I. Vidanovi\'{c}, A. Bala\v{z}, P. Muruganandam and S. K. Adhikari, Comput. Phys. Comm. {\bf 183}, 2021 (2012).





\end{thebibliography}
\end{document}